# FLOW VISUALIZATION OF THE BUOYANCY-INDUCED CONVECTIVE HEAT TRANSFER IN ELECTRONICS COOLING

*Carmine Sapia*

Department of Applied Electronics - University "Roma TRE"
Via della Vasca Navale n. 84, 00146 Rome (Italy)
tel. (+39) 6 5517.7801 - fax (+39) 6 5517.7026 - email: sapia@uniroma3.it

**ABSTRACT**

The aim of this work is to develop a simple optical method for the visualization of the natural convection flow generated in an electronic system during its normal operation. The presented experimental set-up allows to reveal local refractive index changes in a phase objects. A fringe pattern is acquired, through the cooling fluid under analysis, with a digital camera two times: the first one with the fluid at rest, the second one with the thermal load due to the electronic device normal operation. By the means of the MATLAB processing of the acquired images it's possible to reveal the shape and the directions of the thermal flow lines for the cooling fluid. In this way we can obtain a deeper understanding of the optimal convection working volume or information for the optimization of the relative spatial positioning of the several electronic components in a complex electronic system, like a printed circuit board (PCB). The proposed technique has been tested on two typical heat extraction situations recurrent in the electronic devices. In this paper are presented the experimental results of the visualization of the convective flow, in air, for an heat sink and a power resistor.

## 1. INTRODUCTION

The reliability of an electronic system mainly depends on its components working temperature. Thermal management is a key issue in PCB and IC design. Increasing speeds and package densities with reduced feature sizes cause an increasing of power dissipation and higher die temperatures. If the heat generated is not adequately removed, the die temperature increases indefinitely. Failure probability is strictly, exponentially, related to the device operating temperature. The most common kinds of failure are due to increased stresses, to increased electromigration and leakage currents, to differences in thermal expansion caused by the differing coefficient of thermal expansion among materials and to accelerated corrosion mechanism.

Convection, natural or forced, air cooling is still the most utilized method in heat extraction problems in the electronic equipments [1,2]. Natural convection (and radiation) heat transfer, in air, with extended surfaces can manage a thermal load of 0.1 W/m$^2$ with a temperature difference, between device and ambient, of $\Delta T=80$ °C [3,4]. Natural convection cooling, in particular, results simple, cheap and reliable. In compact, low power consumption electronic equipments (as wireless phone, palm handled computers and so on), in sealed system or in solar cell, natural convection is often the only solution available for the cooling technique of the system. It's also the only way to avoid or reduce the acoustic noise (zero dB cooling) related with fan or liquid pumps characterizing other kinds of cooling systems. More complex cooling techniques are necessary only in high power applications. In these cases we have to face with increase of costs and added cooling related space.

In flow visualization problems optical methods are usually preferred [5-9]. Optical techniques are no contact, non destructive kinds of testing. The measure process does not alter the thermal flow field of the cooling fluid under test. Moreover optical methods do not suffer any kind of inertia problems as occurs in other flow visualization techniques which deal with marks particle. Optical methods can visualize thermal phenomenon characterized by very short characteristic time. Non destructive testing, optically implemented, is a powerful investigation tool in thermofluid dynamics problems; it can also validate the results obtained by numerical modelling (CFD) of the heat transfer problem.

## 2. EXPERIMENTAL SET-UP

It's illustrated a simple white-light optical set-up for the monitoring of the refractive index local changes in a fluid: that's what happens in a cooling flow due to the thermal load related to the electronic component normal operation

## 2.1. The principle of the method

Natural convection heat transfer occurs when the convective fluid motion is induced by density differences themselves caused by the heating.

Non uniform refractive index changes in the cooling fluid arise from the density gradients related to the thermal expansion of the fluid. The thermal load of the device causes an optical deflection in the cooling fluid that an opportune light probe can reveal.

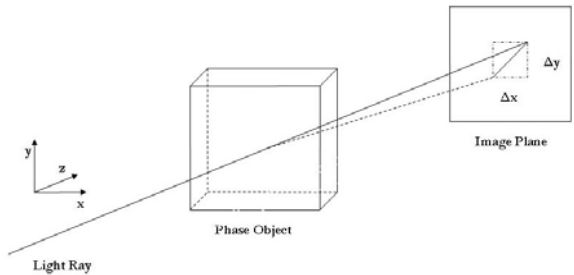

*Figure 1. Light deflection trough an inhomogeneous flow field.*

In figure 1 is summarized the principle of the method. A temperature gradient causes a density variation in a cooling fluid with a consequent local change in the refractive index. The light ray travelling through the fluid is proportionally deflected. Analyzing the deflection of the light probe it is possible to go back to the related temperature gradient. The optical set-up presented in this paper allows to monitor the local variations for the refractive index and, consequently, allows to visualize the convective thermal flow in the cooling fluid.

## 2.1. Experimental procedure

In figure 2 is illustrated a schematic of the experimental set-up developed for the optical flow visualization method utilized.

A step fringe pattern is projected through the electronic device under test and recorded by a CCD camera. The cooling fluid under test is positioned between the reference fringe pattern and the digital camera (*test field area* in figure 2). All the images acquired by a CCD camera (Sony: 768 x 576 x 8bit) are real-time processed by the means of a frame grabber (Data Translation DT-3155). The time sequence of the images is visualized and memorized in a personal computer.

Two different kinds of images are acquired and processed: the first with the working fluid at rest without the thermal load, the second one with the fluid under operative cooling condition. By the means of digital filtering and image processing [10] procedure (MATLAB, release 6.5), it is possible to reveal, qualitatively, the phase gradients and the thermal pattern related to the convective flow emerging from the electronic device.

All the informations for the visualization of the convective flow under test are extracted in white-light without coherent sources. The whole process is digitally carried out.

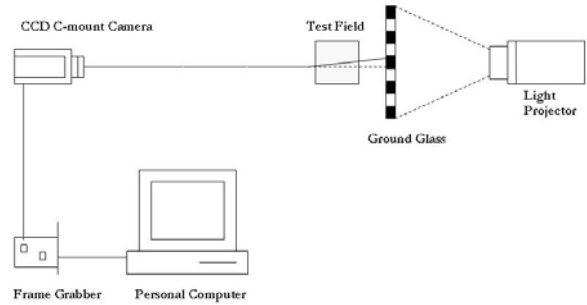

*Figure 2. The experimental set-up.*

## 3. RESULTS

To prove the functionality of the experimental set-up, it was investigated the convective heat transfer of two situations typically recurrent in the electronic systems: a power transistor heat sink and a power resistor. In all the presented situations it was possible to visualize the convective flow generated, in air, by the heated sample and qualitatively understand the shape of the isothermal lines in the test field.

### 3.1. Flow visualization of the buoyancy-induced convective heat transfer, in air, for an heat sink.

In figure 3a - 3c are presented some experimental results for the onset and the development, in air, of the natural convection flow emerging from a power transistor heat sink (ST338K). In an electronic equipment the power transistor is usually positioned in the middle of the heat sink. To reproduce experimentally the described real thermal situation, the structure under investigation (anodized aluminium, 1.5 mm thickness; 41mm x 38mm x 16mm) was centred on a vertical rod electrically heated and controlled in temperature. After an adequate warm-up time (2 hours) of the heater-heat sink system it has been started to acquire images with the optical set-up and the experimental procedure previously described.

The temperature difference, between heat sink and ambient, was settled down at $\Delta T=40$ °C.

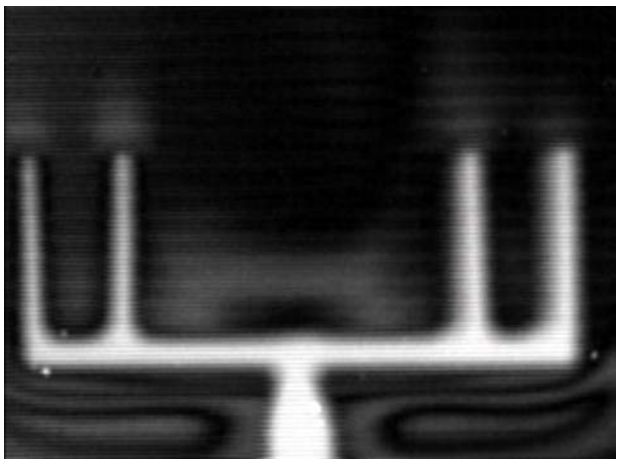

*(a)*

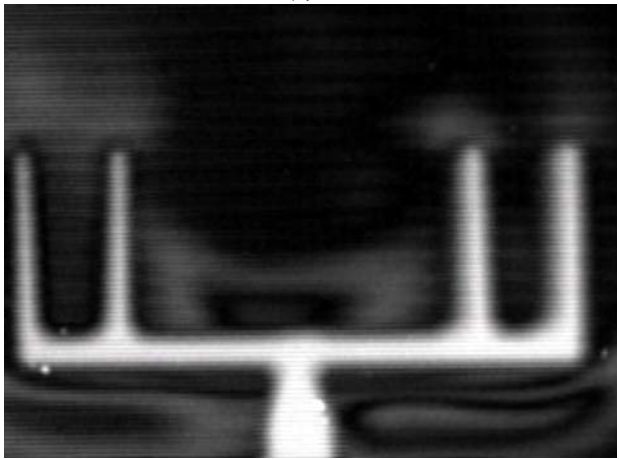

*(b)*

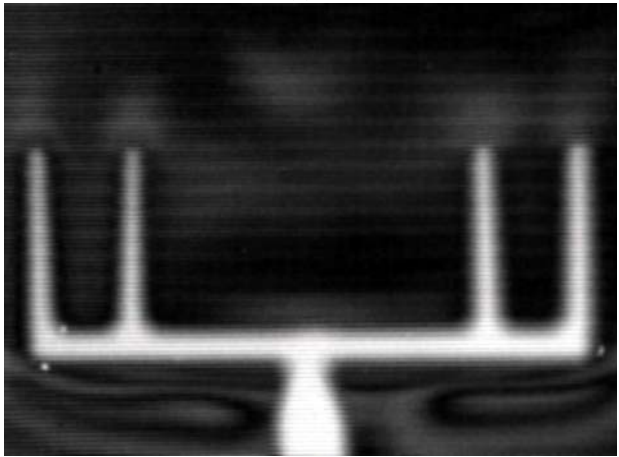

*(c)*

**Figure 3**. *Flow visualization of the buoyancy-induced convective heat transfer, in air, for a power transistor heat sink: (a) the hot spot in the middle causes the onset of a symmetrical thermal bubble; (b) the thermal bubble loses initial symmetry developing into two different lateral plumes; (c) the thermal bubble explodes in various floating separated thermals. At the bottom it's present the onset of the new thermal bubble.*

Non uniform thermal distribution was individuated. As expected, convective flow appears to be stronger in the central part of the heat sink. The main hot spot is clearly highlighted in the middle of the heat sink; hot spots due to spatially not uniform heat flux can cause physical stress and further reduce reliability of an electronic device.

The overheated fluid starts to emerge from the heat sink base with the shape of symmetrical thermal bubble (fig. 3a). The thermal bubble loses initial symmetry developing into two different lateral plumes (fig. 3b). Finally (fig. 3c) the side plumes appear to be detached from the base of the heat sink; the thermal bubble explodes in various floating separated thermals. At the bottom it's also present the onset of the new thermal bubble.

In the figures 3a-3c are also visible the convective plumes at the top of the vertical profiles of the heat sink and lateral thermal bubbles emerging from the base of the narrowest grooves. It's also present a region of compressed overheated fluid below the horizontal base of the heat sink. The hot air emerging from below appears to be compressed by the physical obstruction represented by the base plate of the heat sink. The compressed fluid from below finds only laterally the way to emerge.

The onset and the development of the buoyancy-induced convective heat transfer experimentally observed in this paper is consistent with what reported in literature [11] for the analysis of the natural convection generated by an horizontal hot plate. "Plumes" are defined as extension, usually laminar, of the boundary layer; "thermals" are defined as three-dimensional structures (bubbles) that are detached from the boundary layer floating and emerging in the fluid (fig. 4).

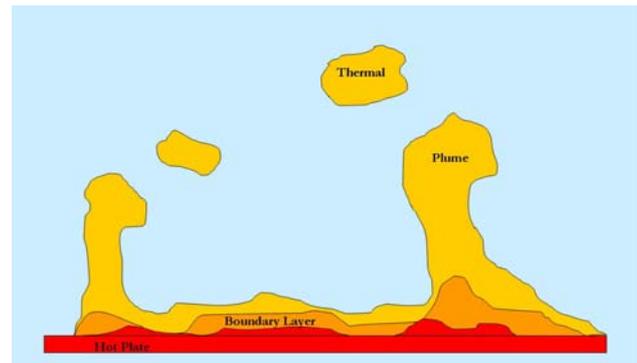

**Figure 4**. *Schematic of thermal and plumes for an horizontal hot plate.*

To enhance the flow visualization results the images obtained with the described optical system have been opportunely post processed, again with the MATLAB

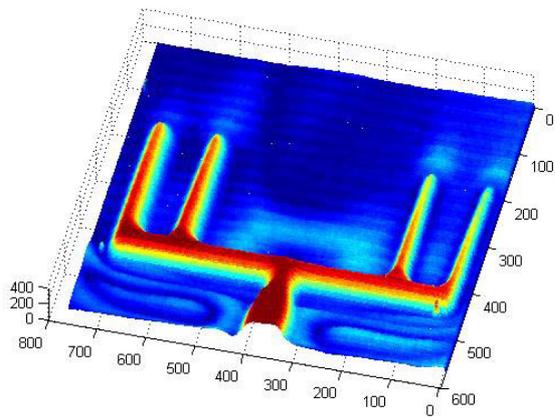

*(a)*

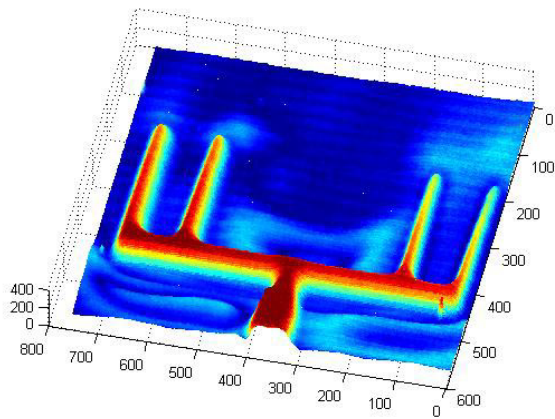

*(b)*

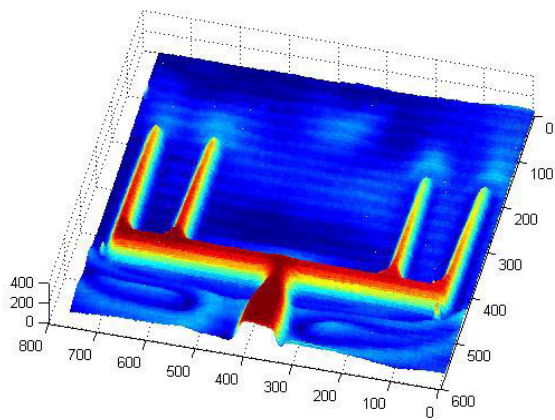

*(c)*

**Figure 5**. *Flow visualization of the buoyancy-induced convective heat transfer, in air, for a power transistor heat sink: the thermal distortion experimentally observed (Fig.3a-3c) for the natural convection air flow is here presented after an height coding.*

software. In figure 5a-5c the thermal distortion experimentally observed for the natural convection air flow is presented after an height coding.

### 3.2. Flow visualization of the convective heat transfer, in air, for a power resistor.

It has been visualized the convective heat flow, in air, for a power resistor (50 W) controlled in temperature with Joule heating. In the figure 6a-6c are presented the flow visualization results. To increase the heat exchange from the resistor to the ambient, the manufacturer has realized the power resistor with a semi spherical extended surface (aluminium; 1 mm thickness). Are also visible the two lateral extensions at the bottom of the device to mount the device on an external metallic plate.

The presented images are obtained with the same procedure and optical set-up described in section 2. The phase gradients related to the thermal expansion of the cooling fluid are presented in three different situations. After an adequate warm up time (2h) the structure of the natural convective flow appears to be completely developed and stabilized (figure 6a). The structure of the flow for the steady state is characterized by two main symmetrical side lobes with an upper plume of heat extraction in the middle. In the bottom of the device is present a bubble of overheated fluid detached from the surface of the resistor.

In figure 6b the shape of the steady state convective flow is disturbed from an added moderate lateral ventilation: the exposed side lobe appears compressed and the upper "chimney" of natural convection results deformed generating a vortex. It's clearly visible some thermal bubbles detached from the surface of the device. As expected in presence of the moderate lateral ventilation the steady state is completely deformed. In this case the thermal refractive index gradients distribution appears to be more complex and distributed involving a larger volume of the cooling fluid.

Increasing the lateral imposed air flow (fig. 6c), the upper plume loses completely his original regularity and direction appearing completely dominated by the side ventilation. It's visible a lateral multi plume structure in the direction of the external air flow with many detached thermal bubbles. The overheated fluid starts to be completely removed from the surface of the power resistor.

### 4. CONCLUSIONS

The experimental work in this paper represents a very simple method for the visualization of non homogeneities in a phase object: the temperature gradients in a cooling

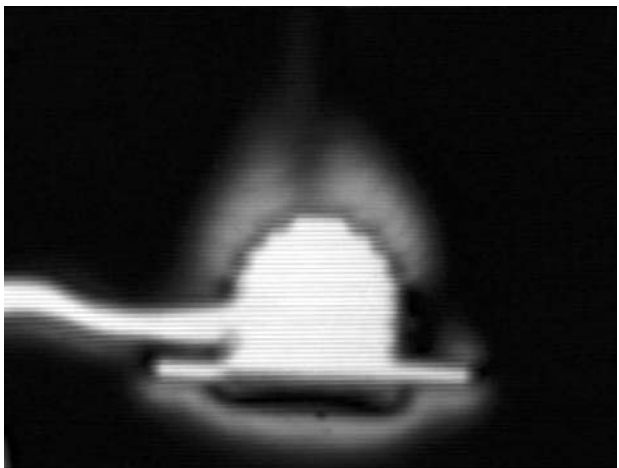

*(a)*

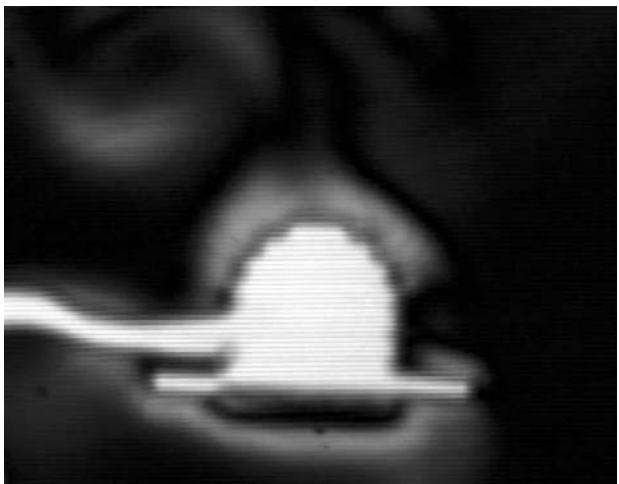

*(b)*

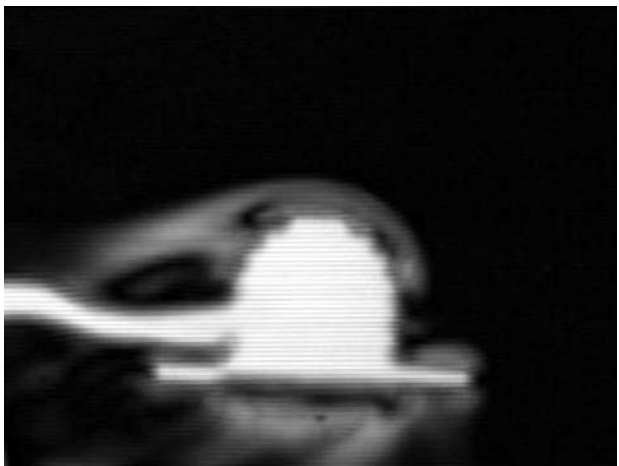

*(c)*

**Figure 6**. *Flow visualization of the buoyancy-induced convective heat transfer, in air, for a power resistor: (a) steady-state flow: two symmetrical side lobes and a main upper plume; (b) moderate overimposed side air flow induces vortex and thermal bubbles for the upper plume; (c) increasing side ventilation, the shape and the direction of the flow is completely dominated by the external air flux.*

fluid for buoyancy-induced convective flow can be visualized.

The set-up was optically implemented: the analysis is absolutely no-contact. The experiment can be carried out without distortion for the thermal flow and, consequently, without perturbation for the heat transfer process and alteration for the temperature gradients in the fluid under test. The system developed was tested on two situations typically recurrent in the electronic systems: an heat sink and a power resistor. In both the cases it was possible to visualize the convective flow generated, in air, by the heated sample and qualitatively understand the shape of the isothermal lines in the test field and the involved volume in the cooling fluid. The results presented show that is possible to monitore the onset and the development of the natural convection thermal flow and the perturbation in the thermal gradient map caused by externally added air flow.

These preliminary results show the potential of the proposed technique to visualize convective flows in electronic cooling problems with a simple and cheap non-invasive optical setup.

The proposed system seems to be suitable only to qualitative analysis but, compared to others well known interferometric optical techniques, the described set-up is less sensitive to environmental instabilities (industrial noise, vibrations); it's possible to make experimental analysis directly *in situ* where the electronic equipment works. Compared to standard Schlieren photography the convective flow field analysis is also carried out without large lenses or mirrors; it's easier suitable to large area whole-field thermal investigations. The results are consistent with other more complex techniques [11].

All the set-up is based on white-light without the need of coherent sources of illumination. The process can also be easy made automatic, easy to use also by non-optically skilled operators.

Experimental methods, for the visualization of the thermal flow and temperature distribution, can provide useful information for the optimization process of the thermal management of an electronic system.

Non destructive optical techniques implemented fully digital can also get the advantage of the recent increasing development of digital acquisition sensors (CCD and CMOS) and image processing techniques.

## 5. REFERENCES

[1] Kim, S.J., Lee, S.W., *Air cooling technology for electronic equipment*, CRC Press, Boca Raton, LA, 1996.

[2] Incropera, F. P., "Convection heat transfer in electronic equipment cooling", *J. Heat Transfer*, vol. 110 No. 4, pp. 1097-1111, 1988.

[3] Cengel, Y., *Heat Transfer*, 2nd ed., McGraw-Hill, New York, 2002.

[4] Incropera, F.P. and DeWitt, D.P., *Fundamentals of Heat and Mass Transfer*, 3rd ed., John Wiley & Sons, New York, 1990

[5] Merzkirch W., *Flow Visualization*, 2nd ed., Academic Press, Orlando, 1987.

[6] Mayinger F., *Optical Measurements*. Springer-Verlag, Berlin, 1994.

[7] Garimella, S. V., "Flow Visualization Methods and their Application in Electronic Systems", *Thermal Measurement in Electronic Cooling*, Azar, K. (ed)., CRC Press LLC, New York, pp. 349-385, 1997.

[8] Merzkirch W.: "Approaches in flow visualization", *Trends in Optical Non-Destructive Testing and Inspection*, P.K. Rastogi and D. Inaudi (ed.). Elsevier, Amsterdam, 2000.

[9] Smith A. J., Lim T. T. *Flow Visualization*, Imperial College Press, London, 2003.

[10] Pratt W. K., *Digital image processing*, Wiley, New York, 1978.

[11] Goldstein R. J., Volino R. J.: "Onset and Development of Natural Convection Above a Suddenly Heated Horizontal Surface", *ASME J. Heat Transfer*, vol. 117, pp. 808-821, 1995.

[12] Aydin, O. and Yang, W-J, "Natural convection in enclosures with localized heating from below and symmetrical cooling from sides", *Int. J. Num. Methods Heat Fluid Flow*, Vol. 10 No. 5, (2000), pp. 519-529.

[13] Ntibarufata, E., Hasnaoui, M., Bilgen, E., Vasseur, P., "Natural convection in partitioned enclosures with localized heating", *Int. J. Num. Meth. Heat Fluid Flow*, Vol. 3, pp. 133-143, 1993.